\documentclass[twocolumn,amsmath,amssymb,floatfix,prb,shownopacs,footinbib,superscriptaddress]{revtex4-1}
\usepackage{amsmath,amssymb,natbib,bm,graphicx,url,epsf,color}
\usepackage[british]{babel}
\usepackage{wasysym}
\usepackage{MnSymbol}

\newcommand{\VQPC}{V_{QPC}}
\newcommand{\VSW}{V_{SW}}

\newcommand{\GQPC}{G_{QPC}}
\newcommand{\Ginf}{G_{\infty}}
\newcommand{\tauinf}{\tau_{\infty}}

\begin{document}

\title{Strong back-action of a linear circuit on a single electronic quantum channel}

%\author{F.D. Parmentier, A. Anthore, S. Jezouin, H. le Sueur, U. Gennser, A. Cavanna, D. Mailly, F. Pierre}
%
%\affiliation{CNRS / Univ Paris Diderot (Sorbonne Paris Cit\'e), Laboratoire de Photonique et de Nanostructures
%(LPN), route de Nozay, 91460 Marcoussis, France}

\author{F.D. Parmentier}
\affiliation{CNRS / Univ Paris Diderot (Sorbonne Paris Cit\'e), Laboratoire de Photonique et de Nanostructures
(LPN), route de Nozay, 91460 Marcoussis, France}
\author{A. Anthore\thanks{anne.anthore@lpn.cnrs.fr}}
\email[e-mail: ]{anne.anthore@lpn.cnrs.fr}
\affiliation{CNRS / Univ Paris Diderot (Sorbonne Paris Cit\'e), Laboratoire de Photonique et de Nanostructures
(LPN), route de Nozay, 91460 Marcoussis, France}
\author{S. Jezouin}
\affiliation{CNRS / Univ Paris Diderot (Sorbonne Paris Cit\'e), Laboratoire de Photonique et de Nanostructures
(LPN), route de Nozay, 91460 Marcoussis, France}
\author{H. le Sueur}
\thanks{Current address: CNRS, Centre de Spectrom\'etrie Nucléaire et de Spectrom\'etrie de Masse (CSNSM), 91405 Orsay Campus, France}
\affiliation{CNRS / Univ Paris Diderot (Sorbonne Paris Cit\'e), Laboratoire de Photonique et de Nanostructures
(LPN), route de Nozay, 91460 Marcoussis, France}
\author{U. Gennser}
\affiliation{CNRS / Univ Paris Diderot (Sorbonne Paris Cit\'e), Laboratoire de Photonique et de Nanostructures
(LPN), route de Nozay, 91460 Marcoussis, France}
\author{A. Cavanna}
\affiliation{CNRS / Univ Paris Diderot (Sorbonne Paris Cit\'e), Laboratoire de Photonique et de Nanostructures
(LPN), route de Nozay, 91460 Marcoussis, France}
\author{D. Mailly}
\affiliation{CNRS / Univ Paris Diderot (Sorbonne Paris Cit\'e), Laboratoire de Photonique et de Nanostructures
(LPN), route de Nozay, 91460 Marcoussis, France}
\author{F. Pierre}
\email[e-mail: ]{frederic.pierre@lpn.cnrs.fr}
\affiliation{CNRS / Univ Paris Diderot (Sorbonne Paris Cit\'e), Laboratoire de Photonique et de Nanostructures
(LPN), route de Nozay, 91460 Marcoussis, France}

\date{\today}

\maketitle

{\sffamily
What are the quantum laws of electricity in mesoscopic circuits? This very fundamental question has also direct implications for the quantum engineering of nanoelectronic devices. Indeed, when a quantum coherent conductor is inserted into a circuit, its transport properties are modified. In particular, its conductance is reduced because of the circuit back-action. This phenomenon, called environmental Coulomb blockade, results from the granularity of charge transfers across the coherent conductor\cite{ingold1992sct}. Although extensively studied for a tunnel junction in a linear circuit\cite{devoret1990dcb,girvin1990dcb,cleland1992dcb,holst1994dcb}, it is only fully understood for arbitrary short coherent conductors in the limit of small circuit impedances and small conductance reduction\cite{yeyati2001dcb,golubev2001dcb,altimiras2007dcb}. Here, we investigate experimentally the strong back-action regime, with a conductance reduction of up to 90\%. This is achieved by embedding a single quantum channel of tunable transmission in an adjustable on-chip circuit of impedance comparable to the resistance quantum $R_K=h/e^2$ at microwave frequencies. The experiment reveals important deviations from calculations performed in the weak back-action framework\cite{yeyati2001dcb,golubev2001dcb}, and matches with recent theoretical results\cite{kindermann2003fcs,safi2004ohmic}. From these measurements, we propose a generalized expression for the conductance of an arbitrary quantum channel embedded in a linear circuit.
}

The transport properties of a coherent conductor depend on the surrounding circuit. First, electronic quantum interferences blend the conductor with its vicinity, resulting in a different coherent conductor (see e.g. ref.~\cite{umbach1987nonlocal}). In addition, the circuit back-action modifies the full counting statistics of charge transfers across coherent conductors\cite{kindermann2003fcs,safi2004ohmic,golubev2005dcb}. This mechanism, which is our concern here, results in violations of the classical impedance composition laws even for distinct circuit elements, separated by more than the electronic phase coherence length. The present experimental work investigates the strong circuit back-action on the conductance of an arbitrary electronic quantum channel.

The circuit back-action originates from the granularity in the transfer of charges across a coherent conductor. Due to Coulomb interactions, an excitation by these current pulses of the circuit electromagnetic modes is possible, which impedes the charge transfers and therefore reduces the conductance of the coherent conductor. This environmental Coulomb blockade is best understood in the limit of a tunnel junction embedded in a circuit of very high series impedance, which is of particular importance for single electron devices\cite{devoret1992set}. In this limit, each time an electron tunnels across the junction, its charge stays a very long time on the capacitor $C$ inherent to the junction's geometry. Consequently, a charging energy $e^2/2C$ has to be paid. Since this energy is not available at low voltages and temperatures, the tunneling of electrons is blocked and the tunnel junction's conductance vanishes. One speaks of `static' Coulomb blockade, because the circuit's dynamical response can be ignored. If now the circuit response time $\tau$ is short enough, the charging energy becomes ill-defined, with an uncertainty $\Delta E\approx h/\tau\gtrsim e^2/2C$. This `dynamical' Coulomb blockade regime corresponds to quantum fluctuations of the charge on the capacitor that are comparable to the electron charge $e$. It is therefore essential to consider the circuit as a quantum object. For a resistor $R$ in series with the tunnel junction, the cross-over between the static and the dynamical Coulomb blockade is at $R\approx R_K\equiv h/e^2\simeq25.8~\mathrm{k}\Omega$. Importantly, the conductance can also be fully suppressed in the dynamical regime, at sufficiently low energy.

The environmental Coulomb blockade was first studied on small, opaque tunnel junctions embedded in linear circuits\cite{devoret1990dcb,girvin1990dcb,cleland1992dcb,holst1994dcb}. The studies were later extended to tunnel junctions of larger conductance\cite{joyez1998dcb} and size\cite{pierre2001dcb}, and to the high frequency domain\cite{hofheinz2011dcb}. To go beyond tunnel junctions, a major theoretical difficulty is that a general coherent conductor, with electronic channels of arbitrary transmission probabilities, cannot be handled as a small perturbation to the circuit. This difficulty was first overcome in the limit of low-impedance linear circuits with a small back-action. In this case, the striking prediction\cite{yeyati2001dcb,golubev2001dcb} and observation\cite{altimiras2007dcb} are that the circuit back-action on the conductance is directly proportional to the amplitude of quantum shot noise in absence of the circuit. However, the even more important and challenging regime of strong back-action remains mostly unexplored and unsolved for arbitrary coherent conductors, despite important advances in that direction\cite{matveev1993tunnel1d,molenkamp1995scalingec,nazarov1999cbwotj,kindermann2003fcs,golubev2005dcb} and a powerful link established with the Luttinger physics of interacting 1D conductors\cite{safi2004ohmic}. The present experimental work investigates this regime on a tunable quantum point contact (QPC) embedded in an on-chip circuit of impedance comparable to $R_K$, beyond reach of perturbative theoretical treatments, resulting in relative reductions of the QPC conductance of up to 90\%.

\begin{figure}[!htbp]
\centering\includegraphics[width=1\columnwidth]{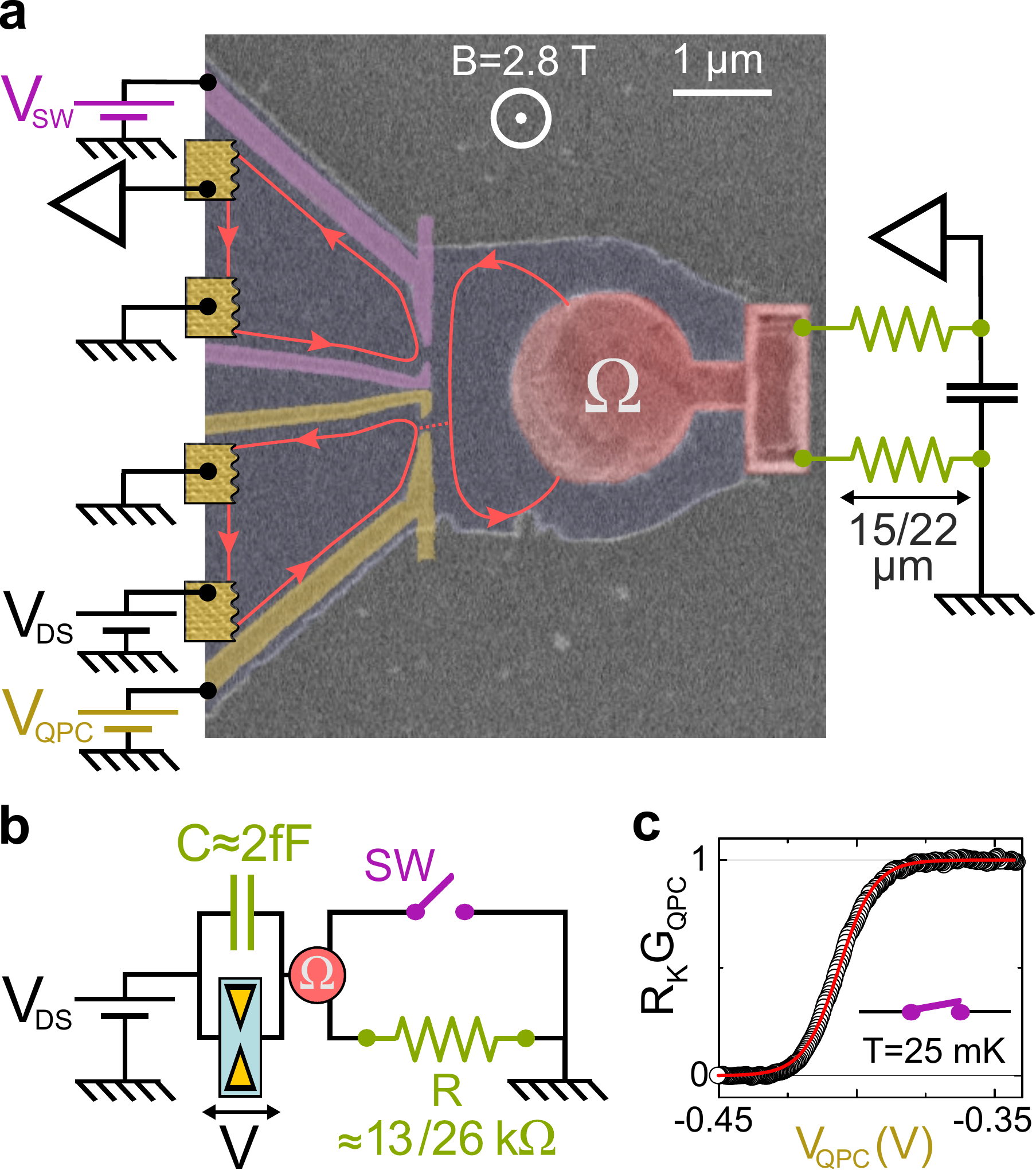}
\caption{
\textbf{Measured devices.} \textbf{a}, SEM micrograph of the $R=26~\mathrm{k}\Omega$ sample. The bottom left metal split gate (yellow) is used to tune the studied QPC. The outer edge channel shown as a red line is partially transmitted at the QPC. A small ohmic contact labeled $\Omega$ (red) is used to connect the 2DEG (light blue) with the series chromium wires symbolized by green resistors. The top left split gate (violet) realizes a switch to short-circuit the on-chip impedance. \textbf{b}, Schematic of the equivalent circuit, with $C$ the parallel geometrical capacitance. \textbf{c}, Conductance $\GQPC$ of the bottom left QPC in \textbf{a} versus the applied gate voltage $V_{QPC}$, for a short-circuited impedance.
}
\label{fig-sample}
\end{figure}

\begin{figure*}[!htb]
\centering\includegraphics[width=0.9\textwidth]{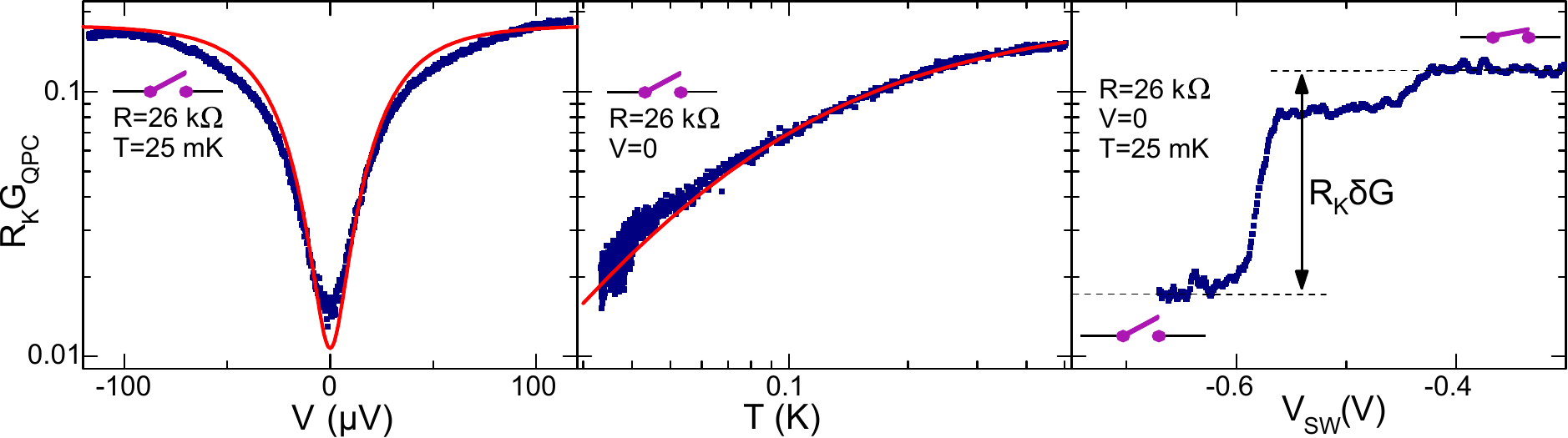}
\caption{\textbf{Back-action signal versus prediction in the tunnel limit}, for the $R=26~\mathrm{k}\Omega$ sample. Predictions (see text) are shown as continuous lines and data as symbols. Left panel, differential conductance $\GQPC$ versus $V$ at $T=25~$mK with the switch open. Center panel, $\GQPC$ versus temperature at $V=0$ with the switch open. Right panel, $\GQPC$ versus switch gate voltage $\VSW$ at $V=0$ and $T=25~$mK. The series resistance, $R=26~\mathrm{k}\Omega$ at $\VSW<-0.6$~V, is short-circuited at $\VSW>-0.4$~V.
}
\label{fig-fig2}
\end{figure*}

The samples are constituted of three basic elements (see Fig.~1a,b): (i) a tunable single electronic channel as a test-bed for coherent conductors, (ii) an on-chip dissipative environment and (iii) a switch to short-circuit the dissipative environment.

(i) We emulate any single-channel short coherent conductor with a tunable QPC formed by field effect in a buried GaAs/Ga(Al)As two-dimensional electron gas (2DEG), using a surface metallic split gate biased at $\VQPC$ (bottom split gate in Fig.~1a). A single-channel short coherent conductor is characterized by the `intrinsic' quantum channel transmission probability $\tauinf\equiv R_K\Ginf\in [0,1]$, with $\Ginf$ the coherent conductor's conductance in the absence of circuit back-action. The single step followed by a well-defined $1/R_K$ plateau of the QPC conductance $\GQPC(\VQPC)$ (symbols in Fig.~1c) shows that the studied QPC can be operated in the single-channel regime, and that its transmission probability can be varied continuously from 0 to 1. The canonical QPC behavior is confirmed by fitting the measured $\GQPC(\VQPC)$ with the standard saddle-point model of a QPC\cite{buttiker1990qpc} (continuous line in Fig.~1c). Note that it is important to break spin degeneracy in order to first study a single electronic channel. Otherwise, the additional channels would partly shunt the surrounding circuit\cite{joyez1998dcb,bagrets2005fcsdcb}. For this purpose, we applied a strong perpendicular magnetic field B=2.8~T corresponding to the integer quantum Hall effect at filling factor 4. Consequently, the current propagates at the edges along four copropagating edge channels. The studied outer edge channel is shown in Fig.~1a as a red line, with an arrow indicating the propagation direction. The three other edge channels (not shown) are always fully reflected at the QPC.

(ii) The second element is the QPC's surrounding circuit, of large dissipative impedance $\mathrm{Re} [Z(\omega)]\sim R_K$ up to microwave frequencies $\omega\sim k_BT/h\sim 1~\mathrm{GHz}$. This is achieved with a nanofabricated on-chip environment modeled by a linear RC circuit in Fig.~1b. The series resistances $R$ is $26~\mathrm{k}\Omega$ for the sample shown in Fig.~1a and $13~\mathrm{k}\Omega$ for a second sample. It is realized by two parallel thin chromium wires of identical lengths $L=22~\mu$m ($15~\mu$m) for $R=26~\mathrm{k}\Omega$ ($13~\mathrm{k}\Omega$) deposited at the surface. These chromium wires can be described as macroscopic linear resistors (see Supplementary Information).
The parallel capacitance $C$ in Fig.~1b corresponds to the shunt capacitor to AC ground of the area delimited by the metal split gates and the series chromium wires. To avoid a capacitive short-circuit of the series resistance at the relevant microwave frequencies, this area must be minimized. For this purpose, the buried 2DEG is connected to the chromium resistors at the surface with a micron-scale AuGeNi ohmic contact (labeled $\Omega$ in Fig.~1a). This micron-scale contact also plays the role of a floating electron reservoir, which breaks the quantum coherence between electrons emitted and arriving at the studied QPC.

(iii) The third element is a switch that allows us to suppress the back-action of the environment by short-circuiting it. This switch is controlled by the voltage $\VSW$ applied to the top split gate in Fig.~1a. A second voltage amplifier (top-left in Fig.~1a) is used to monitor the switch's conductance.

In the present experiment, the reduction $\delta G$ of the QPC conductance $\GQPC$ by the circuit back-action is extracted by three different methods: We measure $\GQPC$ as a function of either the DC voltage $V$ across the QPC (Fig.~2, left panel), the temperature (Fig.~2, center panel), or the gate voltage $\VSW$ controlling the switch (Fig.~2, right panel). In the first two methods, traditionally used to investigate the Coulomb blockade, $\delta G \equiv \GQPC\left(V=0,T\right)-\Ginf$ is obtained by assuming that $\GQPC$ converges toward its `intrinsic' conductance $\Ginf$ for $eV$ or $k_B T$ much larger than $h/RC$ and $e^2/2C$. In the third method, $\Ginf$ is obtained from the QPC conductance measured with a short-circuited environmental impedance. This last, more direct method yields the back-action signal without any particular assumption on its energy dependence, and avoids possible sources of errors related to the transmission energy dependence, sample heating, or the QPC stability over long times.

Figure 2 illustrates the three methods for the same sample of series resistance $R=26~\mathrm{k}\Omega$, and with the QPC set to similar low transmissions. In this near-tunnel limit, the measured voltage and temperature dependence of $\GQPC$, shown as symbols in the left and center panels, can be compared with the known predictions for tunnel junctions\cite{ingold1992sct}. The calculations plotted as continuous lines were performed within the simplified RC model depicted in Fig.~1b. The temperature $T$ was set to that of the dilution fridge mixing chamber, $R=26~\mathrm{k}\Omega$ to the measured value of the on-chip series resistance, and the parallel geometrical capacitance $C$ to the value $C=2$~fF obtained by finite element numerical simulations. The only fit parameter is the transmission in absence of back-action $\tauinf = 0.18$ (left panel) and 0.19 (center panel).
The right panel shows $\GQPC(V=0,T=25~\mathrm{mK})$ versus the voltage $\VSW$ controlling the switch to short-circuit the environment. The capacitive cross-talk between the switch gates and the QPC gates was first calibrated for each sample, then compensated for when sweeping $\VSW$ (see Supplementary Information). For $\VSW<-0.6$~V, the conductance across the switch is zero, and the measured $\GQPC$ corresponds to the conductance reduced by the environmental back-action. As $\VSW$ is increased, the switch's conductance increases in steps corresponding to the successive edge channels transmission. The environmental back-action is found to be suppressed by fully transmitting the two outer edge channels across the switch (see Supplementary Information); the corresponding QPC conductance measured at $\VSW \in [-0.4,-0.3]~$V is taken as $\Ginf$.
We stress that the conductance reductions $\delta G$ obtained from all three methods are consistent with one another, and that we find a good agreement between data and theoretical predictions in the tunnel limit for a known surrounding circuit. This provides strong support for our interpretation of $\delta G$ in terms of environmental back-action. We have now established the experimental principle with a tunnel QPC, and demonstrated the strong back-action regime with a conductance reduction of $90\%$.

\begin{figure}[!htb]
\centering\includegraphics[width=1\columnwidth]{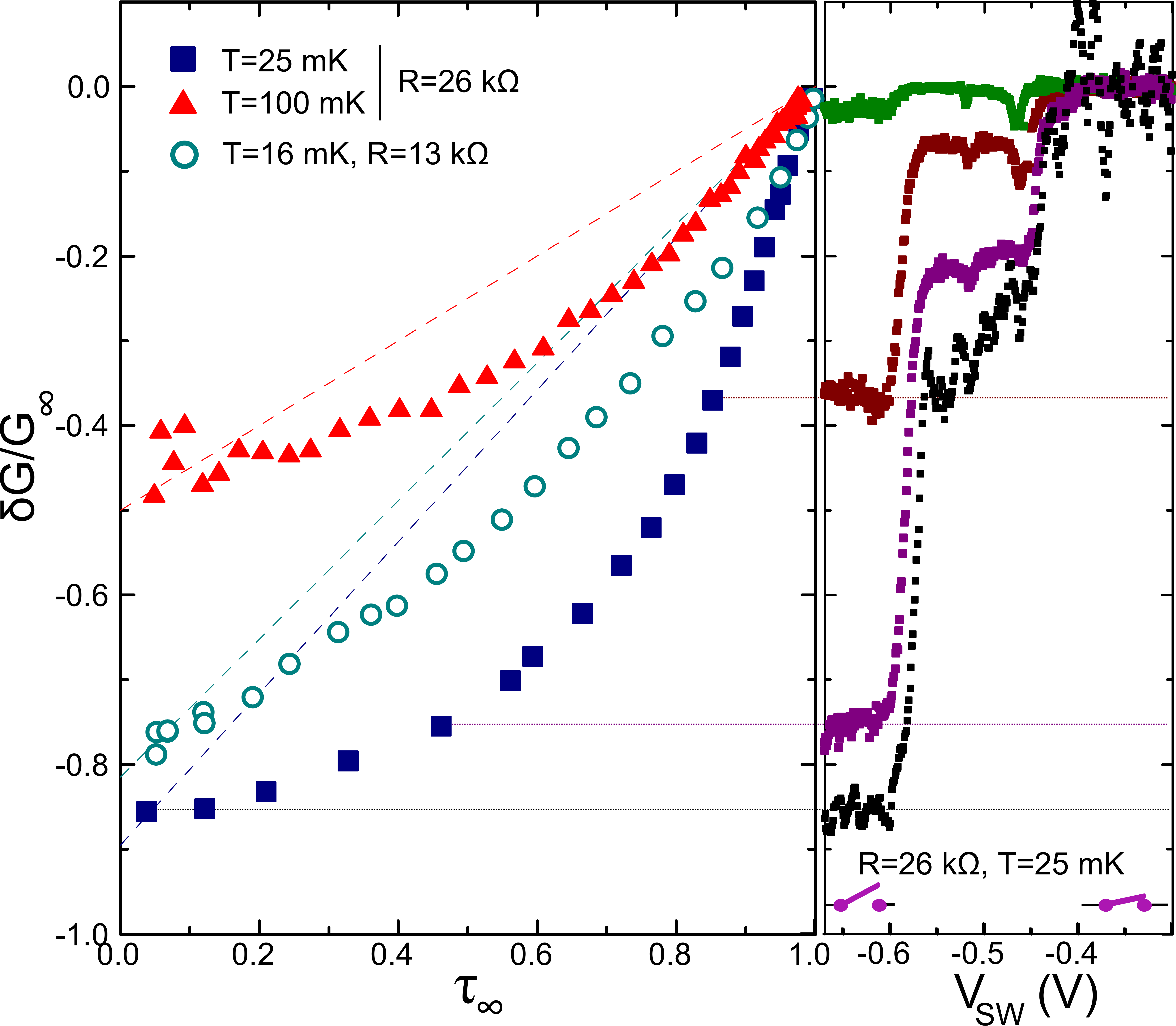}
\caption{\textbf{Environment back-action versus transmission probability.} Left panel, Measured relative back-action amplitude $\delta G/\Ginf$ (symbols) versus `intrinsic' transmission probability $\tauinf$. The data with $R=26~\mathrm{k}\Omega$ are shown for $T=25$~mK ($\blacksquare$) and $T=100$~mK ($\blacktriangle$). Those with $R=13~\mathrm{k}\Omega$ are shown for $T=18$~mK ($\circ$). The dashed lines represent the $(1-\tauinf)$ behavior predicted in the limit of small environmental impedances. Right panel, Sweeps $\delta G/\Ginf(\VSW)$ measured at $\tauinf = \{0.038,0.462,0.853,0.987\}$, respectively from bottom to top, on the $26~\mathrm{k}\Omega$ sample for $T=25$~mK.
}
\label{fig-fig3}
\end{figure}

\begin{figure*}[!htb]
\centering\includegraphics[width=1\textwidth]{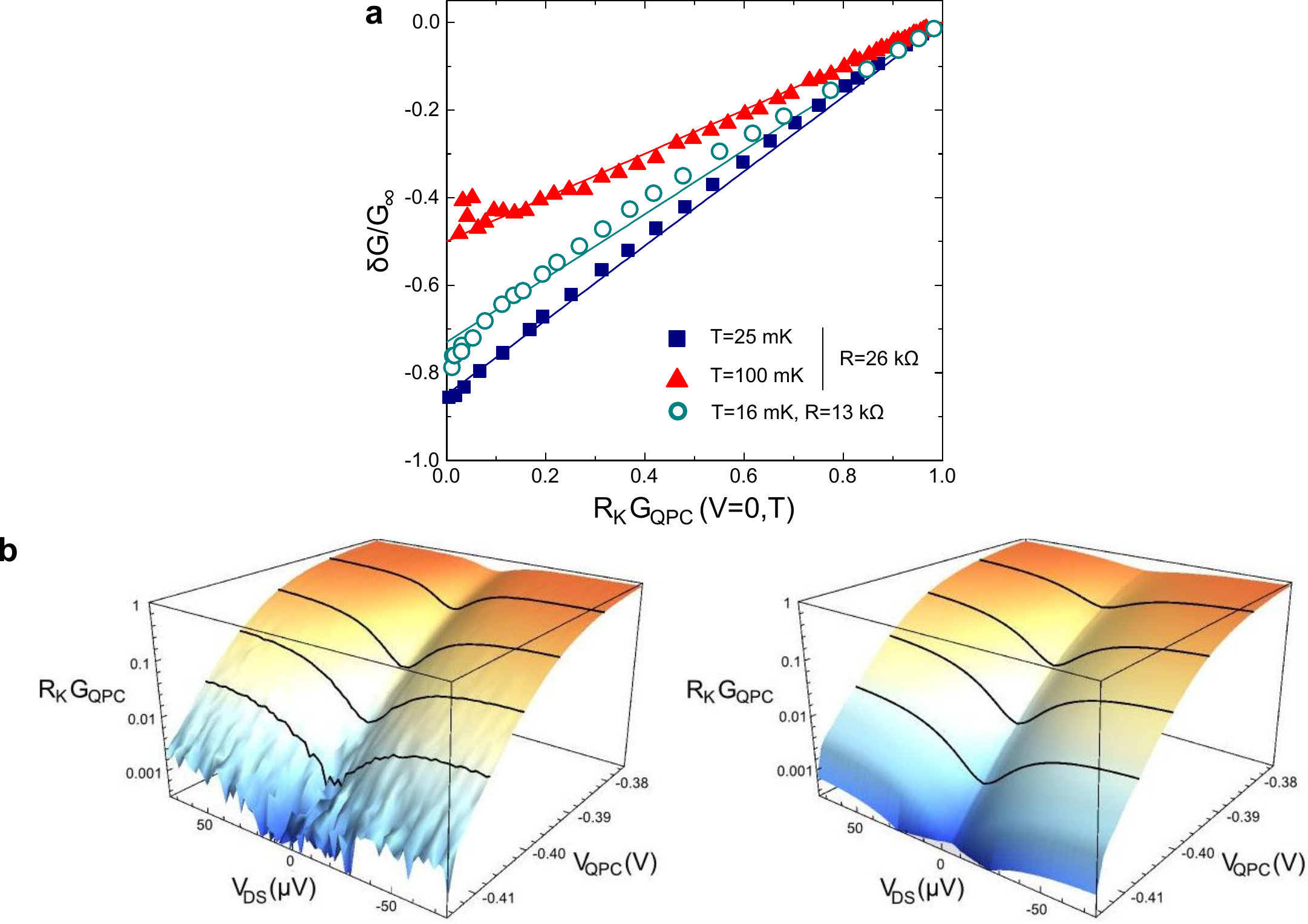}
\caption{\textbf{Comparison between data and extended strong back-action predictions.} \textbf{a}, $\delta G/\Ginf$ (same datasets as Fig.~3) plotted as a function of the conductance $\GQPC (V=0,T)$ in presence of back-action. The straight continuous lines are guides for the eyes. \textbf{b}, Normalized QPC differential conductance $R_K\GQPC$ plotted in Log scale as a function of the QPC gate voltage $\VQPC$ and the applied drain-source voltage $V_{DS}$. The continuous lines correspond to different values of the `intrinsic' transmission probability $\tauinf$, from top to bottom \{0.78,0.37,0.097,0.018\}. Left panel, conductance measured with $R=26~\mathrm{k}\Omega$; right panel, calculations using Eq.~\ref{EqGeneral} with $\tauinf (\VQPC)$ set to the measured $R_K\GQPC (\VQPC,V_{DS}=100~\mu\mathrm{V})$.}
\label{fig-fig4}
\end{figure*}

Next, we investigate the circuit back-action on an arbitrary single-channel coherent conductor characterized by its `intrinsic' transmission probability $\tauinf$. The left panel of Fig.~3 shows as symbols, for both samples, the measured relative reduction of the QPC conductance $\delta G/\Ginf$ due to the circuit back-action when the switch is open, as a function of $\tauinf$. The right panel shows the sweeps $\delta G/\Ginf(\VSW)$ at $\tauinf = \{0.038,0.462,0.853,0.987\}$ for the $R=26~\mathrm{k}\Omega$ sample at 25~mK. First, we observe that $|\delta G/\Ginf|$ is largest in the tunnel limit and diminishes monotonously toward zero as $\tauinf$ increases toward full transmission. However, contrary to predictions and observations in the limit of small environmental impedances\cite{yeyati2001dcb,golubev2001dcb,altimiras2007dcb}, $|\delta G/\Ginf|$ is not proportional to $(1-\tauinf)$ in the full range $\tauinf \in [0,1]$ (dashed lines), but markedly larger at intermediate $\tauinf$. As seen by comparing the data at T=25~mK and 100~mK for the $R=26~\mathrm{k}\Omega$ sample, when the temperature increases, $|\delta G/\Ginf|$ and the deviations to a $(1-\tauinf)$ dependence decrease.

Remarkably, we observe that the back-action correction to the conductance $\delta G/\Ginf$ is instead proportional to $(1-R_K \GQPC(V=0,T))$, for all series resistances and temperatures. This is demonstrated, within experimental uncertainties, in Fig.~4a by plotting $\delta G/\Ginf$ now as a function of $R_K \GQPC(V=0,T)$. This proportionality can be written as $\delta G/\Ginf=(1-R_K\GQPC)E_B$, where $E_B\equiv \lim_{\Ginf\rightarrow 0}\delta G/\Ginf$ is the relative circuit back-action for a small tunnel junction embedded in the same circuit. Using the environmental Coulomb blockade framework for tunnel junctions\cite{ingold1992sct}, $E_B\left( Z,V,T\right)$ can be calculated for arbitrary circuit impedances $Z$, bias voltages $V$ and temperatures $T$. Consequently, solving the above equation for $\GQPC$ allows us to propose a generalized expression for the conductance of a single electronic channel of arbitrary transmission embedded in a linear environment of arbitrary impedance (see also Supplementary Information):
\begin{eqnarray}
\GQPC(V,T)=\frac{\tauinf}{R_K}\frac{1 + E_B\left(Z,V,T \right)}{1+\tauinf  E_B\left( Z,V,T \right)}.
\label{EqGeneral}
\end{eqnarray}

We further tested the proposed equation~1 by comparing in Fig.~4b the measured (left panel) and calculated (right panel) QPC conductance versus the applied bias voltage $V_{DS}$ and the gate voltage $\VQPC$ for $R=26~\mathrm{k}\Omega$. The calculations were performed with Eq.~\ref{EqGeneral} using $\tauinf(\VQPC) \simeq R_K\GQPC(\VQPC,V_{DS}=100~\mu \mathrm{V})$. We find a good agreement between data and theory for $\tauinf < 0.5$, illustrating the validity of this formula even at finite bias voltage. Note that for $\tauinf \gtrsim 0.5$, we find that the measured dip in $R_K\GQPC(V_{DS})$ is significantly narrower than calculations. However, this deviation can be accounted for by including the significant sample heating by the DC current within a simplified model based on the Wiedemann-Franz law (see Supplementary Information).

Moreover, equation~1 agrees with a recent theoretical prediction using a renormalization-group approach\cite{kindermann2003fcs} (see Supplementary Information), and generalizes it to arbitrary impedances, beyond resistances small compared to $R_K$.

Equation~1 could be understood as a direct link between the conductance reduction by the circuit back-action and the quantum shot noise in presence of the circuit. The bridge established for a purely resistive environment between Luttinger physics and environmental Coulomb blockade suggests that $\delta G/\Ginf$ remains proportional to the amplitude of quantum shot noise for arbitrary series impedances\cite{safi2004ohmic}. As pointed out in ref.~\cite{safi2004ohmic}, the quantum shot noise is now strongly modified by the environmental back-action. Although there is no fully developed theoretical framework, the experimental observation $\delta G/\Ginf=(1-R_K\GQPC)E_B$, from which Eq.~1 is derived, would correspond to a quantum shot noise spectral density of the current $S_I$ in presence of back-action that verifies $dS_I/dV=2e\GQPC(1-R_K\GQPC)$.
Significantly, the same expression is verified in absence of circuit back-action\cite{martin1992noise,blanter2000noise}, but using the `intrinsic' transmission probability $\tauinf$ instead of the measured transmission probability $R_K \GQPC$. These relations can be derived exactly in the special case $Z(\omega)=R_K$ (private comm. I.~Safi).

To conclude, we explored the strong back-action of a linear circuit on an arbitrary, single-channel, short coherent conductor. The results suggest the generalized expression Eq.~\ref{EqGeneral} for the environmental back-action, which remains to be derived theoretically. This experiment opens the path for further inquiries of the quantum laws of electricity in nanocircuits. These include the investigation of circuits with coherent conductors in which the environmental back-action can coexist with other phenomena such as the Kondo effect\cite{florens2007dcbkondo}, as well as the investigation of the circuit back-action on the full statistics of charge transfers across a coherent conductor\cite{kindermann2003fcs,safi2004ohmic,golubev2005dcb}.

\subsection*{Methods}
The measurements were performed in a dilution refrigerator down to $T=16$~mK, on two samples tailored in a typical 2DEG. The 2DEG of density $2.5~10^{15}~\mathrm{m}^{-2}$ and mobility $55~\mathrm{m}^2\mathrm{V}^{-1}\mathrm{s}^{-1}$ is buried 94~nm deep in a GaAs/Ga(Al)As heterojunction. The measured differential conductances were obtained by standard lock-in techniques at frequencies below $100~$Hz with custom-made ultra-low noise electronics. In order to measure independently the QPC conductance and the series resistance $R$, and also to test the small ohmic contact, the end of one of the chromium wires realizing $R$ is connected at room temperature to a high impedance voltage amplifier (see Fig.~1a). Due to antenna effects, the impedance of the line toward the amplifier is reduced at the relevant microwave frequencies to about the vacuum impedance $377~\Omega\ll R_K$. This is symbolized in Fig.~1a by a capacitor in parallel with the top-right amplifier. Further details are given in the Supplementary Information.

%\bibliographystyle{nature}
%\bibliography{biblio}

\subsection*{Acknowledgments}
The authors gratefully acknowledge Y.~Nazarov, the Quantronics group and I.~Safi for discussions, F.~Lafont for his contribution to the experiment, and L.~Couraud and C.~Ulysse for their contributions to the nano-fabrication. This work was supported by the ERC (ERC-2010-StG-20091028, \#259033), the ANR (ANR-09-BLAN-0199) and NanoSci-ERA (ANR-06-NSCI-001).

\subsection*{Author contributions}
Experimental work and theoretical analysis: A.A., F.D.P., F.P. and S.J.; nano-fabrication and sample design: A.A., F.P. and H.l.S. with inputs from D.M.; heterojunction growth: A.C. and U.G.; manuscript preparation: A.A., F.D.P., F.P. and U.G. with inputs from coauthors; project planning and supervision: F.P.

\subsection*{Additional information}
Supplementary Information accompanies this paper. Correspondences and requests for materials should be addressed to A.A. and F.P.

\end{document}

\title{Supplementary Information for \\`Strong back-action of a linear circuit on a single electronic quantum channel'}

%\author{F.D. Parmentier, A. Anthore, H. le Sueur, S. Jezouin, U. Gennser, A. Cavanna, D. Mailly, F. Pierre}
%
%\affiliation{CNRS / Univ Paris Diderot (Sorbonne Paris Cit\'e), Laboratoire de Photonique et de Nanostructures
%(LPN), route de Nozay, 91460 Marcoussis, France}

\author{F.D. Parmentier}
\affiliation{CNRS / Univ Paris Diderot (Sorbonne Paris Cit\'e), Laboratoire de Photonique et de Nanostructures
(LPN), route de Nozay, 91460 Marcoussis, France}
\author{A. Anthore\thanks{anne.anthore@lpn.cnrs.fr}}
\email[e-mail: ]{anne.anthore@lpn.cnrs.fr}
\affiliation{CNRS / Univ Paris Diderot (Sorbonne Paris Cit\'e), Laboratoire de Photonique et de Nanostructures
(LPN), route de Nozay, 91460 Marcoussis, France}
\author{S. Jezouin}
\affiliation{CNRS / Univ Paris Diderot (Sorbonne Paris Cit\'e), Laboratoire de Photonique et de Nanostructures
(LPN), route de Nozay, 91460 Marcoussis, France}
\author{H. le Sueur}
\thanks{Current address: CNRS, Centre de Spectrom\'etrie Nucléaire et de Spectrom\'etrie de Masse (CSNSM), 91405 Orsay Campus, France}
\affiliation{CNRS / Univ Paris Diderot (Sorbonne Paris Cit\'e), Laboratoire de Photonique et de Nanostructures
(LPN), route de Nozay, 91460 Marcoussis, France}
\author{U. Gennser}
\affiliation{CNRS / Univ Paris Diderot (Sorbonne Paris Cit\'e), Laboratoire de Photonique et de Nanostructures
(LPN), route de Nozay, 91460 Marcoussis, France}
\author{A. Cavanna}
\affiliation{CNRS / Univ Paris Diderot (Sorbonne Paris Cit\'e), Laboratoire de Photonique et de Nanostructures
(LPN), route de Nozay, 91460 Marcoussis, France}
\author{D. Mailly}
\affiliation{CNRS / Univ Paris Diderot (Sorbonne Paris Cit\'e), Laboratoire de Photonique et de Nanostructures
(LPN), route de Nozay, 91460 Marcoussis, France}
\author{F. Pierre}
\email[e-mail: ]{frederic.pierre@lpn.cnrs.fr}
\affiliation{CNRS / Univ Paris Diderot (Sorbonne Paris Cit\'e), Laboratoire de Photonique et de Nanostructures
(LPN), route de Nozay, 91460 Marcoussis, France}

\maketitle

\section{Supplementary information on methods}

\subsection{Measured samples and experimental techniques}

The samples are nanostructured by standard ebeam lithography in a 94~nm deep GaAs/Ga(Al)As two-dimensional electron gas (2DEG) of density $2.5~10^{15}~\mathrm{m}^{-2}$ and mobility $55~\mathrm{m}^2\mathrm{V}^{-1}\mathrm{s}^{-1}$. The measurements were performed in a dilution refrigerator with a base temperature of $T=16$~mK. All measurement lines were filtered by commercial $\pi$-filters at the top of the cryostat. At low temperature, the lines were carefully filtered and thermalized by arranging them as 1~m long resistive twisted pairs ($300~\Omega /$m) inserted inside 260~$\mu$m inner diameter CuNi tubes, which were tightly wrapped around a copper plate that was screwed to the mixing chamber. The sample was further protected from spurious high energy photons by two shields, both at base temperature. To avoid sample heating, the AC excitation voltages across the sample were smaller than $k_BT/e$.

The differential conductance measurements were performed using standard lock-in techniques at frequencies below 100~Hz. The sample was current biased by a voltage source in series with a $10~$M$\Omega$ or $100~$M$\Omega$ polarization resistance at room temperature. The bias current applied to the drain was converted on-chip into a fixed $V_{DS}$, independent of the QPC conductance, by taking advantage of the well defined quantum Hall resistance to ground of the drain electrode (6.453~k$\Omega$ at filling factor $\nu = 4$). Similarly, the current across the switch is obtained by converting the voltage measured with the top left amplifier in article Figure~1 using the 6.453~k$\Omega$ quantum Hall resistance. The conductances of the QPC, switch and series chromium wires were obtained separately by three point measurements. For the $R=26~\mathrm{k}\Omega$ sample, we used cold grounds directly connected to the mixing chamber of the dilution refrigerator. For the $R=13~\mathrm{k}\Omega$ sample, the ground is at room temperature with a series measurement line resistance of $350~\Omega$.

\begin{figure}[!htph]
\centering\includegraphics[width=1\columnwidth]{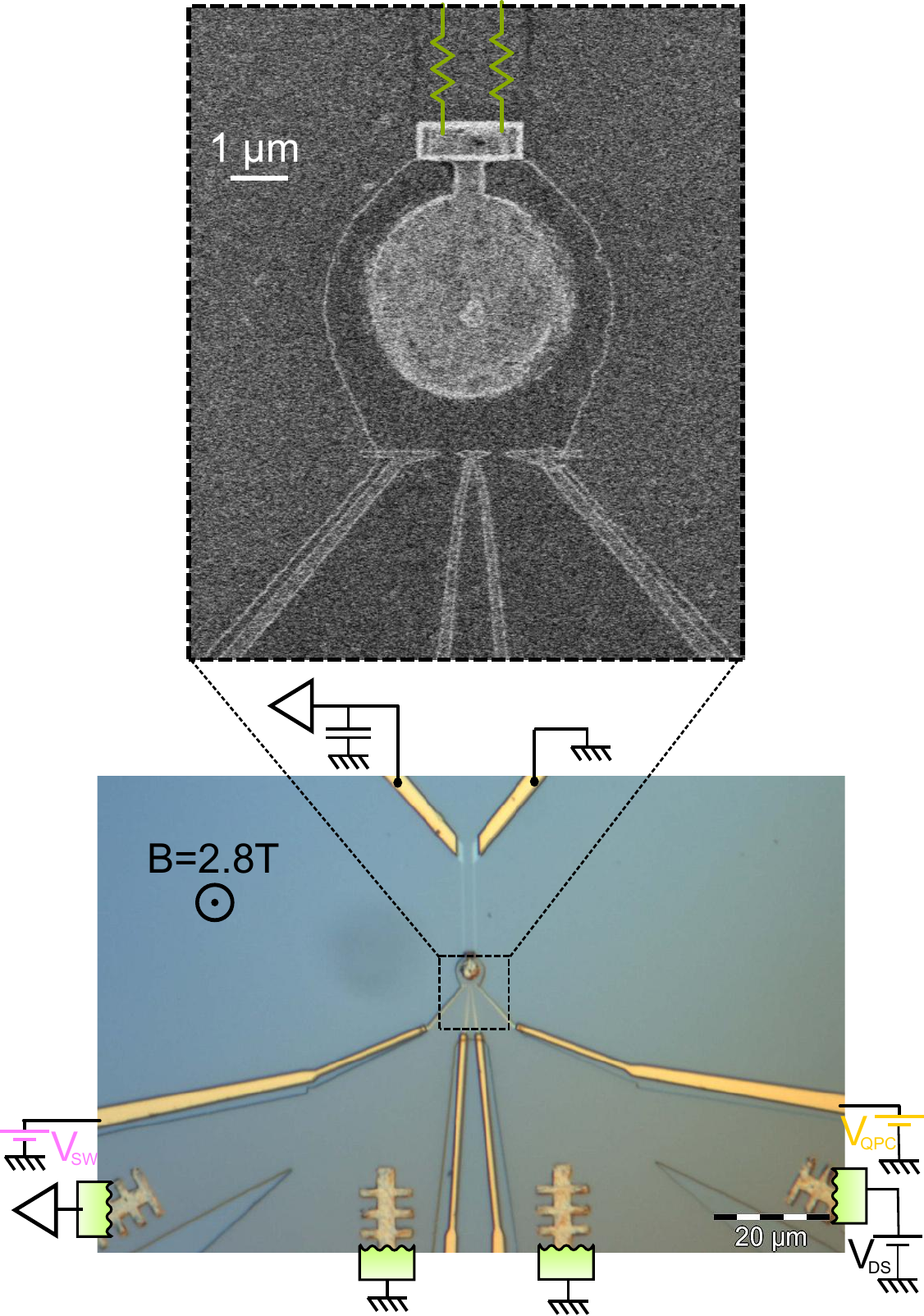}
\caption{\textbf{Sample micrographs} for the $R=13~\mathrm{k}\Omega$ sample. Bottom panel, optical view and schematic of the wiring. Top panel, ebeam micrograph of the central part.
}
\label{fig-SIfig1}
\end{figure}

Micrographs of the $R=13~\mathrm{k}\Omega$ sample are shown in supplementary Figure~1. The larger small ohmic contact in this sample explains the higher computed geometrical capacitance $C=2.3$~fF, instead of $C=2$~fF for the $R=26~\mathrm{k}\Omega$ sample.

\subsection{Supplementary data and discussion regarding the switch operation}

Supplementary Figure 2 shows the same QPC conductance as that displayed in the right panel of article Figure~2, together with the conductance $G_{SW}$ of the switch, which was measured simultaneously.

\begin{figure}[!htph]
\centering\includegraphics[width=1\columnwidth]{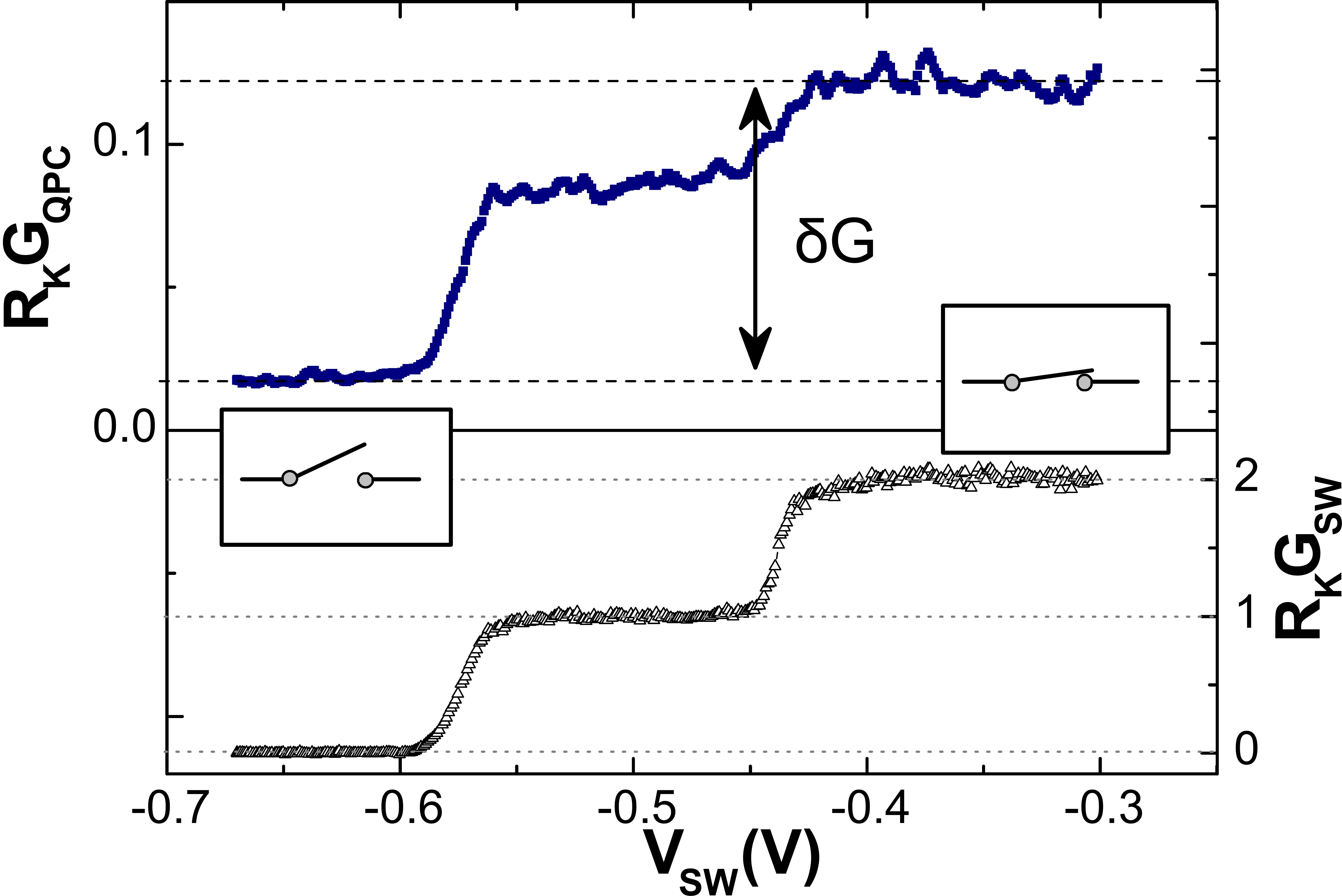}
\caption{\textbf{Switch operation}. Top panel, QPC conductance versus the switch voltage $\VSW$ for the $R=26~\mathrm{k}\Omega$ sample (same data as in right panel of article Figure~2). Bottom panel, conductance across the switch measured simultaneously.
}
\label{fig-SIfig2}
\end{figure}

As $\VSW$ increases above $-0.57$~V, the outer edge channel becomes fully transmitted through the switch and $\GQPC$ simultaneously undergoes a large step up. In a simple edge channels picture, one expects that the environmental back-action is now completely short-circuited. Indeed, the outer edge channel, emitted by the small ohmic contact and transmitting the environmental back-action, is diverted from the studied QPC toward a grounded ohmic contact. Surprisingly, we observe another, smaller step up in the QPC conductance as a second edge channel is transmitted through the switch at $\VSW \sim -0.4~$V. This could be related to $e^2/2C \sim 40~\mu$eV being non-negligible compared to the Zeeman splitting $E_Z\approx 70~\mu$eV between the first two edge channels. For $\VSW \in [-0.4,-0.3]~$V, the two outer edge channels are fully transmitted across the switch and the QPC displays a fixed conductance taken as $\Ginf$. Note that $\GQPC$ does not increase significantly when the third edge channel becomes fully transmitted (data not shown), indicating the circuit back-action on the outer edge channel is essentially short-circuited when $G_{SW}=2/R_K$.

To further demonstrate that the environmental back-action is suppressed with the switch set to transmit fully the two outer edge channels, we show in supplementary Figure~3 the surface plots of the measured $\GQPC(V_{DS},\VQPC)$ for the $R=26~\mathrm{k}\Omega$ sample with the two outer edge channel transmitted across the switch (switch closed, left panel) and with the switch open (right panel, same data as article Figure~4b). As expected if the series impedance is fully short-circuited, there are no signatures of environmental back-action in the drain-source voltage dependence of $\GQPC$ when the switch is closed.

\begin{figure}[!htph]
\centering\includegraphics[width=1\columnwidth]{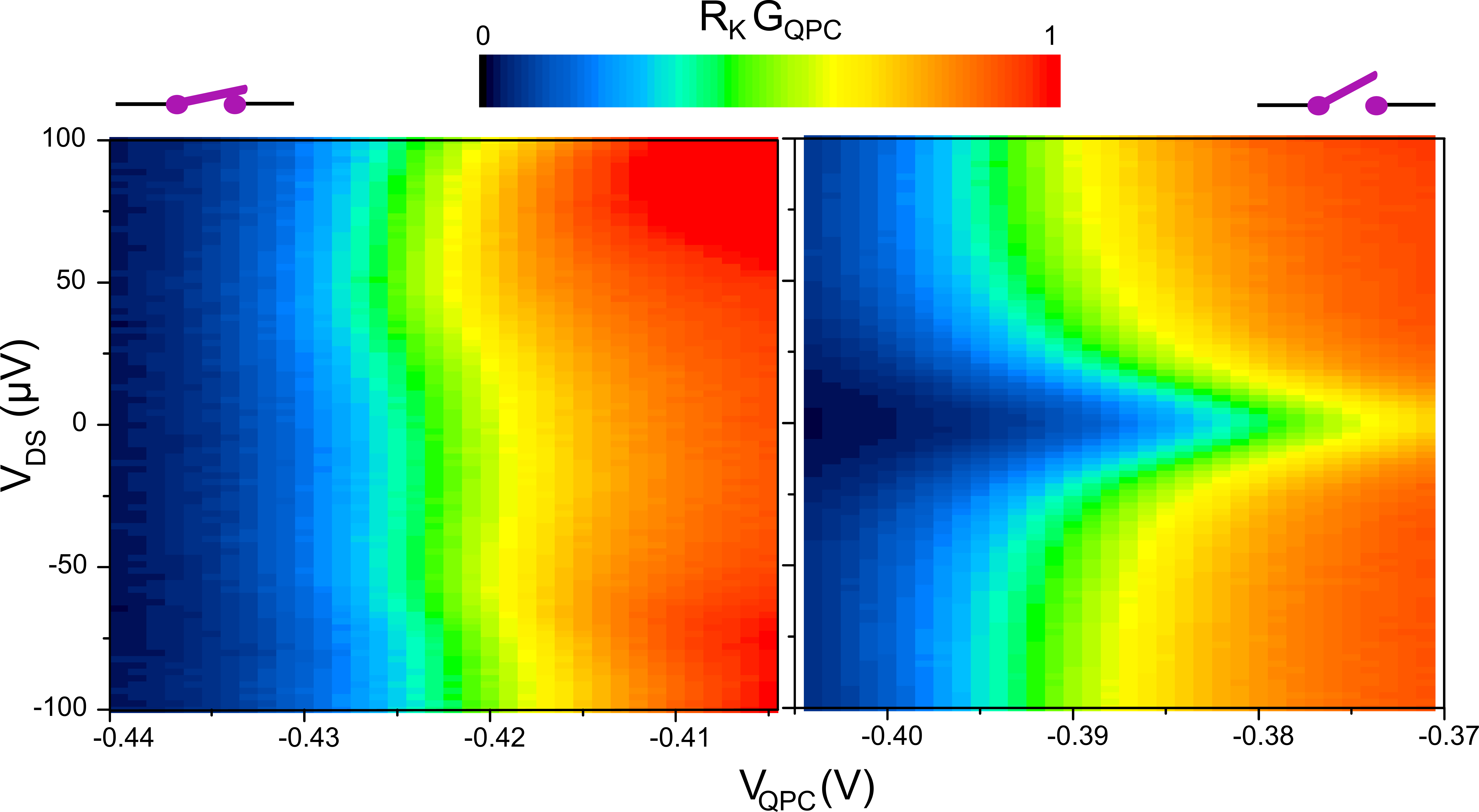}
\caption{\textbf{Switching ON and OFF the voltage bias dependence of $\GQPC$.} $\GQPC(V_{DS},\VQPC)$ measured on the $R=26~\mathrm{k}\Omega$ sample for the two outer edge channels fully transmitted across the switch (switch closed, left panel) and for a switch conductance set to zero (switch open, right panel).
}
\label{fig-SIfig3}
\end{figure}

Importantly, it is necessary to correct for the capacitive cross-talk between the top metal gate controlling the switch and the studied QPC. This capacitive cross-talk is compensated for by changing simultaneously the voltage $\VQPC$ by an amount proportional to $\VSW$. This coefficient is adjusted to a fixed value 0.1 by assuring that the QPC conductance does not change when sweeping $\VSW$ along each plateau $G_{SW}=0$ (resistive environment) and $G_{SW}=2/R_K$ (short-circuited environment), see supplementary Figure~2. The value of this proportionality coefficient is then fixed for all temperatures and $\Ginf$. The quality of the cross-talk correction can be checked on the right panel of article Figure~3 from the flatness of the $\GQPC(\VSW)$ plateaus with the switch fully open ($\VSW<-0.6~$V) and closed ($\VSW>-0.4~$V).

\subsection{Tests performed on the small ohmic contact}

We have checked the quality of the electrical connection between the small ohmic contact and the buried 2D electron gas.

The test was performed with both the conductance of the studied QPC and the conductance of the switch in the middle of the very large and robust $G=2/R_K$ plateau. Assuming that the two outer (inner) edge channels are fully transmitted (reflected) across the studied QPC and the switch, we find for both samples that the reflection of each of the two outer edge channels on the small ohmic contact is smaller than $0.01$. Thus, we consider that the small ohmic contact is perfectly connected to the 2D electron gas.

\subsection{Tests performed on the on-chip chromium resistors}

The on-chip chromium resistors are described as linear macroscopic resistors. We performed several checks to validate this description.

First, this is justified if the phase coherence length $L_\phi$ of electrons in the chromium wires is much smaller than the wire length, $L=15$ and $22~\mu m$ respectively for the $R=13~\mathrm{k}\Omega$ and $R=26~\mathrm{k}\Omega$ samples. An upper bound for the phase coherence length is obtained from the absence of detectable universal conductance fluctuations, whose rms amplitude $\delta G_\mathrm{rms}$ remained smaller than $0.0003/R_K$ at $T=60$~mK for the two wires in series of the $R=13~\mathrm{k}\Omega$ sample. The amplitude of universal conductance fluctuations gives access to the phase coherence length through
\begin{equation}
L_\phi = \frac{3 \pi}{8}\frac{L^3}{L_T^2}(R_K \delta G_\mathrm{rms})^2,
\end{equation}
with $L_T=\sqrt{D \hbar / k_B T}$ the thermal length and $D$ the diffusion coefficient. Injecting in this equation $D\approx 10^{-4}~\mathrm{m^{2}s^{-1}}$, obtained from the chromium density of states at the Fermi energy $\nu\approx 3 \times 10^{47}~\mathrm{J^{-1}m^{-3}}$, and from the wire geometry and resistance, we find $L_\phi\lesssim 200~$nm. Therefore, $L_\phi$ is at least two orders of magnitude smaller than the wire lengths.

Second, the resistances of the chromium wires were checked to be unaffected by the circuit back-action, in contrast to a coherent diffusive conductor. This check was done by monitoring the resistance of one chromium wire while short-circuiting its surrounding circuit, here made of the second chromium wire in parallel with the two QPCs.

Last, we checked that the voltage and temperature dependence of the chromium resistances remained small, below $2\%$ variations in the explored range.

\subsection{Remark regarding the (absence of) effect of interactions between co-propagating edge channels}

It was shown recently that the co-propagating quantum Hall edge channels interact strongly one with another \cite{lesueur2010relaxiqhr}. It is legitimate to wonder whether this effect could play a role in the present investigation of the circuit back-action. We can argue it is not the case for three main reasons:

First, in most cases we extract the effect of the circuit back-action on the QPC conductance by short-circuiting the circuit, without changing the quantum Hall physics. This allows us to separate the circuit back-action from possible interaction effects between edge channels.

Second, experimentally, we find little voltage dependence on the QPC conductance when the short-circuit is closed, as seen on Supplementary Figure~3. This shows that if there is an effect of inter edge channel interactions on the QPC conductance, it remains small compared to the circuit back-action.

Third, theory predicts that even if two co-propagating edge channels interact strongly, this would not result in a voltage dependence of the conductance of a QPC \cite{levkivskyi2008dmzinu2,giamarchi2003qp1d}. This was verified experimentally in the same sample where strong inter edge channels interaction was evidenced \cite{altimiras2010nesiqhr,lesueur2010relaxiqhr,altimiras2010ctrlrelaxqhr}.

\subsection{Toolkit for dynamical Coulomb blockade calculations in the tunnel regime}

Calculations of the conductance of a coherent conductor in the tunnel limit (with very small transmission probabilities for all the electronic quantum channels) in presence of environmental back-action were made with the efficient formulation of the dynamical Coulomb blockade theory for small tunnel junctions given in ref.~6. In this section, we recapitulate the used expressions.

The differential conductance $G\equiv dI/dV$ of a small tunnel junction embedded in an electromagnetic environment of impedance $Z(\omega)$, at temperature $T$, and for a voltage $V$ across the junction reads:
\begin{multline}
G(V,T) \equiv \frac{E_B\left(Z,V,T \right)+1}{R_T} =  \frac{1}{R_T} \Bigg[ 1  + 2 \int_0^{+\infty} \pi t\left(\frac{k_B T}{\hbar}\right)^2\\
\times \mathrm{Im}\left[e^{J(t)}\right] \cos \frac{e V t}{\hbar} \sinh^{-2} \frac{\pi t k_B T}{\hbar} dt \Bigg],
\end{multline}
with $R_T\equiv 1/\Ginf$ the tunnel resistance assumed to be very large compared to the environmental impedance $\mathrm{Re}[Z(\omega)] \ll R_T$.

For the simplified RC model of the electromagnetic environment shown in article Figure~1b ($Z(\omega)=R/(1+iRC\omega)$), $J(t)$ reads:
\begin{multline}
J(t)=\frac{\pi R}{R_K}\Bigg( \left(1-e^{- \mid t\mid/RC} \right) \left( \mathrm{cot}\frac{\hbar}{2RCk_BT}-i \right)-\frac{2  k_B T \mid t \mid}{\hbar} \\ + 2 \sum_{n=1}^{+\infty}\frac{ 1-e^{-\omega_n \mid t\mid}}{2 \pi n \left[ (RC\omega_n)^2-1\right]}   \Bigg),
\end{multline}
with $\omega_n=\frac{2\pi n k_B T}{\hbar}$ the Matsubara's frequencies $n$ being an integer and
\begin{multline}
2 \sum_{n=1}^{+\infty}\frac{ 1-e^{-\omega_n  t}}{2 \pi n \left[ (RC\omega_n)^2-1\right]} \\= - \frac{1}{\pi}\big[2\gamma +\Psi(-x)+\Psi(x)+2\ln(1-y)+ \\ \frac{y}{1+x}~_2 F_1(1,1+x,2+x,y)+ \frac{y}{1-x} ~_2F_1(1,1-x,2-x,y) \big],
\end{multline}
where $\gamma$ is here Euler's constant, $\Psi$ is the logarithmic derivative
of the Gamma function, $~_2F_1$ is the hypergeometric
function, $y=\mathrm{exp}(\frac{-2 \pi t k_B T}{\hbar})$, and $x=E_C R_K/(2\pi^2R k_B T)$ with $E_C=e^2/(2C)$, the charging energy.

\section{Supplementary discussion}

\subsection{Supplementary discussion regarding the theoretical predictions of Kindermann and Nazarov}

Kindermann and Nazarov \cite{kindermann2003fcs} extended to large back-action corrections the theoretical predictions for an arbitrary short coherent conductor embedded in a purely dissipative circuit characterized by a resistance $R$ very small compared to $R_K$. In this section, we recall the theoretical prediction of ref.~7 for the conductance of a single channel coherent conductor and compare it to the experimental findings.

Kindermann and Nazarov predict that the transmission probability $\tau(E)$ of an electronic quantum channel in series with a resistance $R\ll R_K$ at zero temperature varies with the energy $E$ like:
\begin{equation}
\frac{d\tau(E)}{d\ln(E)}=\frac{2R}{R_K} \tau(E) (1-\tau(E)).
\label{EqKN}
\end{equation}

In order to calculate the relative conductance reduction this expression must be complemented by a high energy cutoff, typically of the order of the charging energy $E_C$. Using a cutoff at precisely $E_C$, supplementary Equation~\ref{EqKN} can be integrated (see Eq.~13 in ref.~7):
\begin{equation}
\tau(E)=\tauinf \frac{ \left(\frac{E}{E_C}\right)^{2R/R_K}}{1+ \tauinf \left( \left(\frac{E}{E_C}\right)^{2R/R_K}-1\right)}.
\label{EqKNint}
\end{equation}
The above expression is equivalent to article Equation~1, which expresses formally the experimental findings, with the substitutions $\tau(E)/R_K\rightarrow \GQPC(V,T)$. Indeed $\left(\frac{E}{E_C}\right)^{2R/R_K}$ corresponds, up to a prefactor of order 1, to the function $\left(E_B(Z,V,T)+1\right)$ for the impedance $Z$ corresponding to the series resistance $R$ and the parallel capacitance $C$ at zero temperature $T=0$ (see ref.~8).

A more direct comparison with the data can be made by recasting supplementary Equation~\ref{EqKNint} into:
\begin{equation}
\frac{\tau(E)}{\tauinf}-1=(1-\tau(E))\times \Bigg(\left(\frac{E}{E_C}\right)^{2R/R_K}-1\Bigg).
\label{EqKNrecast}
\end{equation}
With the above formulation, one immediately sees that the predicted relative back-action amplitude (left-hand side) is proportional to $(1-\tau(E))=(1-R_K\GQPC)$, as was observed experimentally.

\subsection{Alternative formulation of manuscript Equation~1}

Equation~1 of the manuscript can be recast in an expression that better emphasizes the scaling laws discussed by Kindermann and Nazarov in ref.~7 (private comm. Y.V.~Nazarov):
\begin{equation}
\frac{\tau_1/(1-\tau_1)}{\tau_2/(1-\tau_2)} = \frac{E_B(Z_1,V_1,T_1)+1}{E_B(Z_2,V_2,T_2)+1},
\label{EqONErecast}
\end{equation}
where $\tau_{1(2)}\equiv R_K\GQPC(Z_{1(2)},V_{1(2)},T_{1(2)})$. This formulation holds for a given electronic quantum channel and any two sets of parameters $\{Z_1,V_1,T_1\}$ and $\{Z_2,V_2,T_2\}$. Remarkably, the two sets of parameters are here connected directly, without the intervention of the `intrinsic' transmission $\tauinf$, and also for different electromagnetic environments $Z_1$ and $Z_2$.

\subsection{Supplementary discussion on heating effect at finite bias}

This section deals with the deviations observed between the data and the prediction of article Equation~1, at intermediate voltage bias for transmissions $\tauinf\gtrsim 0.5$. The discrepancy data-calculation in the shape of $\GQPC(V_{DS})$ is explained if we take into account heating effects as detailed in this section.

At low temperature phonons are inefficient to evacuate from the electronic fluid the heat injected locally. Furthermore, the presence of series resistors comparable to the resistance of the QPC impedes the electronic heat currents toward the cold reservoirs (drain, source, AC ground). Consequently, there are no easy escape paths for the injected Joule power. This results in an increase of the electronic temperature that is not taken into account in article Equation~1. In this section, we use a simplified model to estimate the electronic temperature and compare the calculations using this increased temperature to both the data and the prediction without heating.

\begin{figure}[!htph]
\centering\includegraphics[width=1\columnwidth]{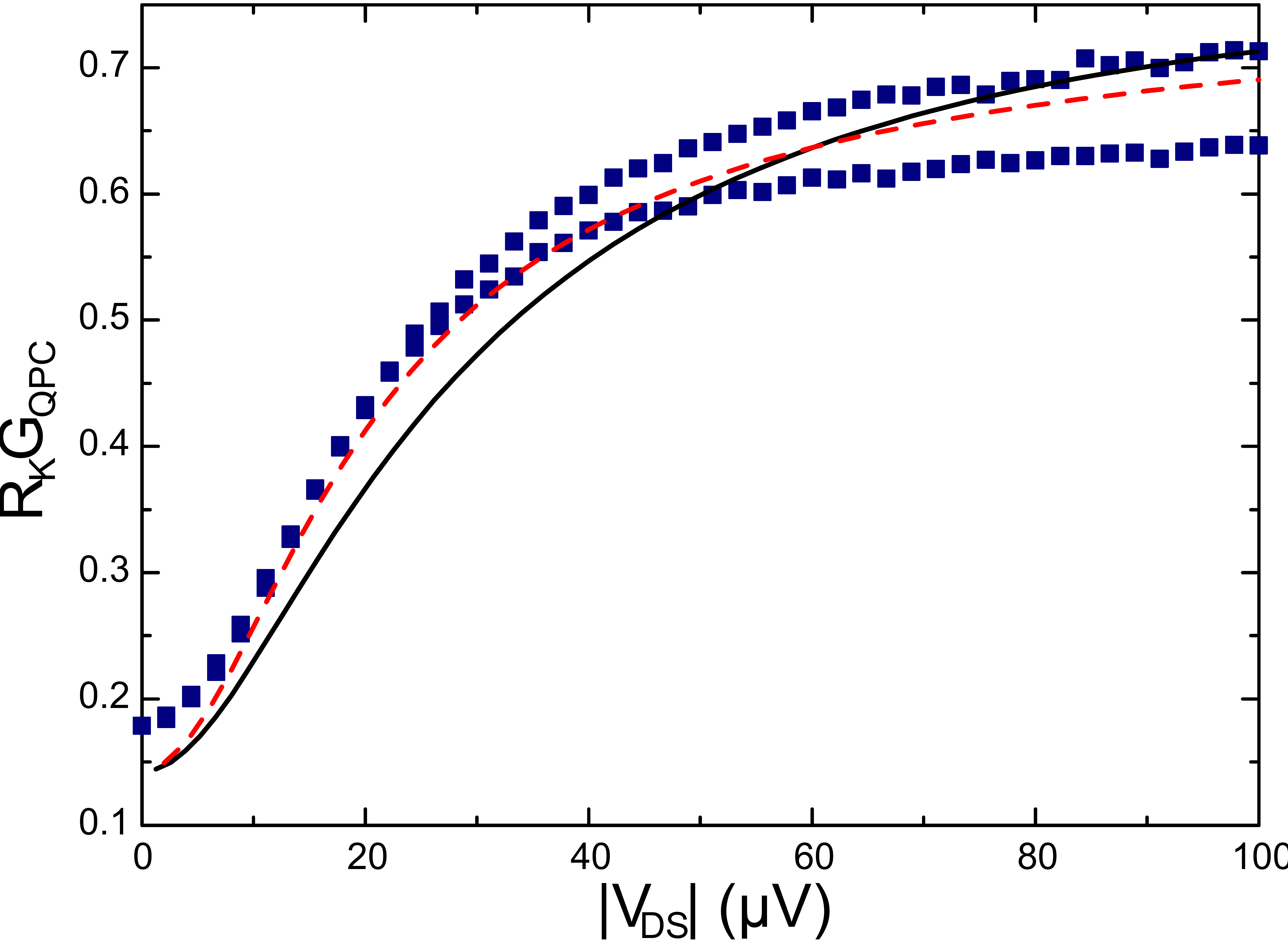}
\caption{\textbf{Effect of heating at finite bias.} Blue symbols: measured $R_K \GQPC$ for $\tauinf=0.77$ on the $R=26~\mathrm{k}\Omega$ sample, and plotted versus the applied $|V_{DS}|$. These data are extracted from article Figure~4b, with the two branches corresponding to positive and negative $V_{DS}$. Black continuous line: calculated $\GQPC=\frac{dI}{dV}$ without taking into account heating effect, using $R=26~\mathrm{k}\Omega$, $C=2fF$ and $T=25~mK$. Red dashed line: calculated $\GQPC$ including heating effects with our simplified model (see text).
}
\label{fig-SIfig4}
\end{figure}

\subsubsection{Description of the model}

Our simplified model of heating relies on the Wiedemann-Franz law. The main hypotheses to calculate the environmental back-action including heating are:\\
\textbf{H1:} We ignore heat dissipation outside the electronic degrees of freedom. In particular, we ignore the dissipation toward phonons.\\
\textbf{H2:} We calculate the electronic heat currents using the Wiedemann Franz law with the average electrical conductance $G_\mathrm{avg}=I/V$ in presence of the back-action.\\
\textbf{H3:} We use for the temperature of the electromagnetic environment the calculated average electronic temperature in the chromium wires.\\

\subsubsection{Temperature calculations}

The hypothesis H1 implies that the Joule power injected into the micron-scaled ohmic contact is evacuated by the outgoing electronic heat currents across the QPC ($J_{QPC}$) and the series resistance ($J_R$):
\begin{equation}
G_\mathrm{avg}V^2/2+(V_{DS}-V)^2/(4R)=J_{QPC}+J_R.
\end{equation}
Note that the Joule powers $VI$ and $(V_{DS}-V)I$ are equally distributed on both sides of the considered conductor, which explains a factor $1/2$. Note also that the resistance taken into account for the injected Joule power is $2R$ since the end of only one of the two chromium wires is connected to ground.

The hypothesis H2 gives the outgoing heat currents as a function of the micron-scale ohmic contact temperature $T_\Omega$ and the mixing chamber temperature $T$:
\begin{align}
J_{QPC}&=\frac{\pi^2}{6 h} k_B^2(T_\Omega^2-T^2)R_K G_\mathrm{avg},\\
J_{R}&=\frac{\pi^2}{6 h} k_B^2(T_\Omega^2-T^2)R_K/R.
\end{align}

Solving this set of equations gives:
\begin{align}
T_\Omega=\sqrt{T^2+\frac{3 e^2V_{DS}^2}{\pi^2k_B^2}\frac{R G_\mathrm{avg} }{ (1 + R G_\mathrm{avg})(1+2 R G_\mathrm{avg})}}.
\end{align}

Similarly, the heat equation is solved along the chromium resistors to compute the spatial average of the chromium resistors' temperature $T_R$. The calculation of $\GQPC(V)$ is then performed with article Equation~1. Note that the function $E_B(Z,V,T)$ is now obtained using $T_\Omega$ on one side of the QPC and $T$ on the other side, whereas the environment temperature is set to $T_R$. Although in such non-equilibrium situation the efficient formulation of $E_B(Z,V,T)$ recapitulated in the above section `Toolkit for dynamical Coulomb blockade calculations in the tunnel regime' does not work, the calculation can be performed with the standard formulation \cite{ingold1992sct}. Note also that because the temperatures depend on $G_\mathrm{avg}$, several iterations must be performed to reach a stable solution.

The effect of heating, calculated within this simplified model, on the environmental back-action is illustrated in supplementary Figure~4 at $\tauinf=0.77$, close to maximum heating for a given $V_{DS}$. The observed good agreement with the data strongly suggests that in this range of conductances and voltage bias the sample heating has to be taken into account.

%\bibliographystyle{nature}
%\bibliography{biblio}